# Realization of bulk insulating property and carrier manipulation in reversible spin current regime of the ideal topological insulator TlBiSe$_2$

Kenta Kuroda[1, †, *], Gaku Eguchi[2], Kaito Shirai[1], Masashi Shiraishi[2], Mao Ye[3], Koji Miyamoto[3], Taichi Okuda[3], Shigenori Ueda[4], Masashi Arita[3], Hirofumi Namatame[3], Masaki Taniguchi[1, 3], Yoshifumi Ueda[5], Akio Kimura[1, *]

[1]*Graduate School of Science, Hiroshima University, 1-3-1 Kagamiyama, Higashi-Hiroshima 739-8526, Japan.*

[2]*Department of Electronic Science and Engineering, Kyoto University, Katsura, Nishikyo-ku, Kyoto 615-8510, Japan.*

[3]*Hiroshima Synchrotron Radiation Center, Hiroshima University, 2-313 Kagamiyama, Higashi-Hiroshima 739-0046, Japan.*

[4]*Synchrotron X-ray Station at SPring-8, National Institute for Materials Science, 1-1-1 Kouto, Sayo, Hyogo 679-5148, Japan.*

[5]*Kure National College of Technology, Agaminami 2-2-11, Kure 737-8506, Japan.*

[†] *Present address: Fachbereich Physik und Zentrum für Materialwissenschaften, Philipps-Universität Marburg, Germany*

## Abstract

The surfaces of three-dimensional topological insulators (TIs) characterized by a spin-helical Dirac fermion provide a fertile ground for realizing exotic phenomena as well as having potential for wide-ranging applications. To realize most of their special properties, the Dirac point (DP) is required to be located near the Fermi energy with a bulk insulating property while it is hardly achieved in most of the discovered TIs. It has been recently found that TlBiSe$_2$ features an in-gap DP, where upper and lower parts of surface Dirac cone are both utilized. Nevertheless, investigations of the surface transport properties of this material are limited due to the lack of bulk insulating characteristics. Here, we present the first realization of bulk insulating property by tuning the composition of Tl$_{1-x}$Bi$_{1+x}$Se$_{2-\delta}$ without introducing guest atoms that can bring the novel properties into the reality. This result promises to shed light on new exotic topological phenomena on the surface.

After the discovery of three-dimensional topological insulators (TIs), intensive studies of the surface Dirac fermions, characterized by massless energy dispersions with a spin-helical texture in $k$ space, have appeared (1, 2). Fascinatingly, when the Dirac point (DP) is located near the Fermi energy ($E_F$) within the bulk energy gap, the novel surface state has a great potential to lead electromagnetic topological phenomena (1-5).

Up to now, a number of materials that exhibit a surface Dirac cone with a spin-helical texture have been studied by angle-resolved photoemission spectroscopy (ARPES) and spin-resolved ARPES (SARPES) (6-18). Spin-helical textures were clearly identified above the DP in tetradymite compounds, such as $Bi_2Se_3$ and $Bi_2Te_3$, but the relevant feature was obscured below it (15, 16). This probably stems from their particular band structures, in which the lower part of the surface Dirac cone is buried in the bulk valence band (BVB), as is actually observed in the ARPES spectra (7, 8) (Fig. 1 **A**). Indeed, the electrical control of n-type to p-type surface Dirac fermions is a current important subject. It is, however, insufficient to realize a surface-dominated conduction due to the contribution of undesirable bulk conduction and therefore many efforts have been paid to reduce the bulk conduction in these compounds (19-21).

Recently, a single surface Dirac cone has been experimentally verified for $TlBiSe_2$, which adopts a rhombohedral crystal structure without a van der Waals gap (10-12). Of note is the size of the bulk band gap energy, which is estimated to be 300 meV, is comparable to that of $Bi_2Se_3$. Furthermore, as shown in Fig. 1**B**, the DP of the surface state has been found to be energetically well isolated from the BVB and bulk conduction band (BCB). This situation is not realized in $Bi_2Se_3$ (6, 9) as seen in Fig. 1**A**. Owing to the in-gap DP feature, the spin-helical texture was identified not only for the upper part of surface Dirac cone but also for the lower part in the previous SARPES study (22). In addition to these talented electronic structures, a massive surface Dirac cone has been observed in this system even if time reversal symmetry is preserved without introducing any magnetic dopants (23). The occurrence of these features at the surface of $TlBiSe_2$ motivates us to further explore this surface to possibly realize macroscopic topological phenomena. However, despite the potential advantages of this material, the study of surface transport properties is hindered by the bulk metallic conduction due to spontaneous carrier doping effects (10-12). These difficulties prevent us from investigating topological phenomena and thus the production of a bulk insulating state in this material is strongly required.

Here, we show how to control the bulk carrier density in $TlBiSe_2$ and lead the material to a bulk insulating state in the topological transport regime. Our SARPES measurements combined with variable light polarizations and photon energies have revealed the novel spin-helical texture in the surface Dirac cone. Chemical composition analysis has revealed that several kinds of defects contribute to a spontaneous electron doping. On the basis of this result, we synthesized a high-quality crystal of $Tl_{1-x}Bi_{1+x}Se_{2-\delta}$ (26), which enables us to tune the bulk carrier density and manipulate the Dirac point energy ($E_{DP}$) below and above $E_F$. Finally, a bulk insulating behavior, that is, a negative temperature coefficient of the longitudinal resistivity ($\rho_{xx}$) has been achieved accompanying the metallic surface conduction at low temperature. Our key finding, that the ideal topological insulating property can be achieved without any guest atoms, unambiguously offers a platform for a wide range of technical applications.

Figure 1**C** shows the constant energy contours (CECs) near the Γ point from $E_F$ to a binding energy ($E_B$) of 600 meV for the sample with $x$=0.064 showing a clear Dirac-cone like energy dispersion. One can see that the DP is located at $E_B$=310 meV, which is slightly lower than that reported in previous works (10-12), probably because of surface carrier doping induced by gas absorption after sample cleaving (26). As Fig. 1**D** shows, the Fermi surface of the surface state exhibits isotropic feature, in contrast to the situation in $Bi_2Te_3$ and $Bi_2Se_3$ (7, 8). The absence of an anisotropic feature for the surface Dirac cone, the so-called warping effect, may indicate that electron scattering can be suppressed as long as time reversal symmetry is preserved (27, 28). Here, we note that the surface Dirac cone can still be clearly identified in Fig. 1**E**, even below the DP, which owes to the characteristic band structure where the DP is energetically isolated from the BVB.

In order to perform the SARPES measurement, we accurately determined $k$ positions from the spin-integrated ARPES result with the higher momentum resolution (26). Figure 1**F** shows the spin-resolved energy distribution curves (EDCs) acquired with linearly polarized light (horizontal). Here, the spin-up and spin-down spectra are plotted with upright and inverted triangles, respectively. During the spin-resolved measurement, there was a slight energy shift of the DP by 20 meV towards $E_F$, which is, however, ignorable on the energy scale used for the spin- resolved measurement (26). In the normal emission configuration ($k_∥$=0 Å$^{-1}$), the spin-up and spin-down spectra are identical. At $θ$=-1.8 ° ($k_∥$~-0.06 Å$^{-1}$), the spin-up state is found to be centered at $E_B$=150 meV, while the spectral weight of the spin-down state is comparatively flat. In going from $θ$=-1.8 ° to +1.8 °, the spin-up state shifts to higher $E_B$ with increasing $θ$. The spin-down feature shows similar behavior, but the direction of the energy shift is opposite. The measured spin polarizations of the surface Dirac cone at several emission angles are shown in Fig. 1**G**. Here, it should be emphasized that the spin polarizations are clearly resolved not only in the upper part of surface Dirac cone, but also in the lower part, and furthermore, the spin polarizations are inverted with respect to the Γ point, which is consistent with the previous SARPES work on the same material (22). Recent theoretical and experimental investigations on spin polarizations of a surface Dirac cone indicate that final state effects play a role, that is, the observed spin-helical state by means of photoelectron spectroscopy is sensitive to the incident light polarization and experimental geometry (24, 25). Therefore, to evaluate the surface state spin polarizations more accurately, SARPES measurements with variable light polarizations and photon energies are important. To confirm this reversed spin polarization feature, the spin-resolved measurement has also been performed with circularly polarized light. The observed spin polarization with circularly polarized light at +1.8 ° shows the same feature as that seen for linear polarization in Fig. 1**G**. Note that the spin polarizations are also independent of the incident photon energy (see supplementary information, SI.5 of Ref. 26). The determined spin orientations are denoted with the arrows in Figs. 1**D** and 1**E**, which signify a spin-helical texture in $k$ space, as depicted in Fig. 1**H**.

Before attempting a bulk carrier control, we have explored the chemical composition of grown samples with an electron probe micro-analysis (EPMA) to understand the origin of the spontaneous electron doping effect, and found that it is non-stoichiometric. This result indicates the occurrence of several kinds of defects. Accordingly, we have synthesized single crystalline $Tl_{1-x}Bi_{1+x}Se_{2-\delta}$ (TBS, hereafter) with manipulating atomic compositions, where $x$ and $\delta$ indicate the amount of substituted atoms in cation and anion-site defects, respectively. In this work, we have focused on a control of the bulk carrier density through $x$ values instead of $\delta$ as described in supplementary material (see SI. 2 of Ref. 26). The evolution of surface Dirac cone for TBS with different $x$ is shown in Fig. 2**A**-**D**. By taking the DP energy ($E_{DP}$) position as a reference, the $E_F$ shifts downwards by 100 meV for TBS with $x=0.032$ with respect to that for $x=0.064$. For lower $x$ values, TBS goes into a bulk insulator phase, where the $E_F$ is located inside the bulk energy gap, and the DP is still present below $E_F$. For TBS with $x=0.015$, the $E_F$ undergoes a further downward shift and the DP is pushed above $E_F$. To quantitatively evaluate the $E_F$ shift of the surface state as a function of $x$, the band dispersions need to be compared among various compounds with different $x$. Figure 2**E** shows the result of peak plots obtained from EDCs and momentum distribution curves (MDCs). Since the band dispersions in all TBS systems of this work are identical to each other when their energy positions are plotted with respect to the $E_{DP}$, the observed energy shift is found to follow a rigid band-like picture, despite a total energy shift as large as ~0.4 eV. To confirm the gapless feature of the surface state, the EDC at the center of the surface Brillouin zone is depicted for TBS with $x=0.032$ in Fig. 2**F**. Note that the EDC is composed of a single peak at the DP that can be reproduced well with a single Lorentzian peak, which signifies the absence of energy gap and its massless feature even after a change of compositions, in sharp contrast to the gapped surface state found for $TlBi(Se_{1-x}S_x)_2$ (23).

In order to experimentally examine if the $E_F$ shift is the bulk origin, a deeply buried bulk electronic state has been investigated by hard X-ray photoelectron spectroscopy (HAXPES) using a photon energy of 5.95 keV, where the inelastic mean-free-path of the photoelectron reaches as large as ~8 nm (29). Figure 3**A** shows the Se *3d* core-level spectra for compounds with different $x$ values, in which the peaks are found around $E_B$ of 52 ~ 55eV. One can see that the observed peaks shift to lower $E_B$ with decreasing $x$. Similar energy shifts are also found for the other core-levels. Typical results for the Tl *5d*, Bi *5d* and Se *3d* core-levels are plotted in Fig. 3**B** together with the $E_{DP}$ shift determined by the ARPES measurements shown in Fig. 2**A**-**D**. One can see that all the bulk core-levels show a monotonic energy shift to lower $E_B$ with decreasing $x$ and the maximum energy shift of ~330 meV is obtained at the minimum $x$ in this study. This result indicates that all the observed core-level shifts are attributable to the bulk $E_F$ shift caused by the suppression of spontaneous electron doping.

By referring to our previous ARPES study for the characteristic energy band structure of TBS system (10), the bottom of BCB is located around $E_B=220$ meV for TBS with $x=0.064$. We show the relationship between the BCB bottom and the $E_F$ in the inset of Fig. 3**B** (26). For TBS with $x=0.042$, the bottom of BCB is pushed up but still crosses the $E_F$ and furthermore the $E_{DP}$ shift coincides with the $E_F$ shift in the bulk. However, the DP drastically shifts upward by ~300 meV for TBS with $x < 0.032$, where the $E_F$ goes into the bulk energy gap region. The observed difference in the $x$-dependent energy shift for the bulk and surface might be ascribed to the emergence of the band bending due to the different bulk insulating level (30).

The bulk insulating behavior accompanying the metallic surface conduction has been finally confirmed by the transport measurements. Figure 3**C** shows the $\rho_{xx}$ and the Hall coefficient ($R_H$) measured as a function of temperature for TBS with $x$=0.025. As a remarkable finding here, the $\rho_{xx}$ is found to show a negative temperature coefficient as a hallmark of bulk insulating behavior. Note that the slope of the temperature coefficient becomes smaller below 40 K and the $\rho_{xx}$ shows a "kink" at this temperature (dashed line). To get more insight into the origin of the observed "kink" in the resistivity, the $R_H$ is plotted versus temperature as shown in the inset. We find that the |$R_H$| decreases with elevating temperature above 40K that is typical for insulators as is well described with the Arrhenius law (dashed line). Then it turns into constant below ~40 K as typically observed in metals. This result signifies that the undesirable bulk conduction is well suppressed in low temperature, and the surface metallic channel contributes significantly to the whole conduction, which is in excellent agreement with the ARPES result (see Fig. 2**C**).

The realization of the bulk insulating property in this material gives us great advantages for future applications because it enable us to manipulate the $E_F$ with a gate voltage of an ambipolar transistor (19-21). In this respect, with tuning the $E_F$ above and below the DP, one can control even the spin current on the surface with n-type and p-type carrier transitions. In particular, in the case of this material, the reversal spin current property is expected without the bulk conduction thanks to the in-gap DP feature. We should note that our results are unique because either guest atom doping or fabrication of multinary compounds is usually required to control bulk carriers for naturally electron-doped TIs (31-33). However, these methods might lead to considerable lattice disorder that can disturb electron transport, thus substantially reducing the carrier mobility. Our methodology, which avoids these techniques, provides a path to a new style of material design of TIs with high surface electron mobility.

In conclusion, we have established a methodology to realize a bulk insulating property in the ternary topological insulator TBS system. We have found unequivocal evidence for an initial state spin texture that reverses its helicity above and below the DP at the surface using our spin- and angle-resolved photoemission spectrometer. Further, we have found that the natural electron doping is caused not only by anion site defects but is also due to defects at the cation sites. Starting from this disadvantageous situation, we have demonstrated that the DP of the functional surface state can be manipulated below and above $E_F$ by controlling atomic compositions close to stoichiometric values without introducing any foreign elements. The observed temperature dependences of $\rho_{xx}$ and $R_H$ have unambiguously proved the bulk insulating property with the metallic surface conduction. Our result has the advantage of being able to obtain an isolated DP from bulk continuum states in this material, which allows access to both of the spin-helical textures without any bulk carrier interruptions. These findings forge a new direction for topological insulators towards realizing an ambipolar gate control with high mobility electrons for future spintronic devices, and also provide an ideal platform to study new exotic phenomena.

**References and Notes:**

1. M. Z. Hasan, C. L. Kane, Topological Insulators. *Rev. Mod. Phys.* **82**, 3045-3067 (2010).

2. X-L. Qi, S-C. Zhang, Topological insulators and superconductors. *Rev. Mod. Phys.* **83**, 1057-1110 (2011).

3. R. Li, J. Wang, X-L. Qi, S-C. Zhang, Dynamical axion field in topologicalmagnetic insulators. *Nature Phys.* **6**, 284–288 (2010).

4. R. Yu *et al.*, Quantized anomalous Hall effect in magnetic topological insulators. *Science* **329**, 61–64 (2010).

5. K. Nomura, N. Nagaosa, Surface-Quantized Anomalous Hall Current and the Magnetoelectric Effect in Magnetically Disordered Topological insulators. *Phys. Rev. Lett.* **106**, 166802 (2011).

6. Y. Xia *et al.*, Observation of a large-gap topological insulator class with a single Dirac cone on the surface. *Nature Physics*. **5**, 398-402 (2009).

7. Y-L. Chen *et al.* Experimental Realization of a Tree-Dimensional Topological Insulator, $Bi_2Te_3$. *Science* **325**, 178-181 (2009).

8. K. Kuroda *et al.*, Hexagonally Deformed Fermi Surface of the 3D Topological Insulator $Bi_2Se_3$. *Phys. Rev. Lett.* **105**, 076802 (2010).

9. S. Kim *et al.*, Surface Scattering via Bulk Continuum State in the 3D Topological Insulator $Bi_2Se_3$. *Phys. Rev. Lett.* **107**, 056803 (2011).

10. K. Kuroda *et al.*, Experimental Realization of a Three-Dimensional Topological Insulator Phase in Ternary Chalcogenide $TlBiSe_2$. *Phys. Rev. Lett.* **105**, 146801 (2010).

11. T. Sato *et al.*, Direct Evidence for the Dirac-Cone Topological Surface State in the Ternary Chalcogenide $TlBiSe_2$. *Phys. Rev. Lett.* **105**, 136802 (2010).

12. Y-L. Chen *et al.*, Single Dirac Cone Topological Surface State and Unusual Thermoelectric Property of Compounds from a New Topological Insulator Family. *Phys. Rev. Lett.* **105**, 266401 (2010).

13. S. Souma *et al.*, Topological Surface States in Lead-Based Ternary Telluride $Pb(Bi_{1-x}Sb_x)_2Te_4$. *Phys. Rev. Lett.*, **108**, 116801 (2012).

14. K. Kuroda *et al.*, Experimental Verification of $PbBi_2Te_4$ as a 3D Topological Insulator. *Phys. Rev. Lett.* **108**, 206803 (2012).

15. Z.-H. Pan *et al.*, Electronic Structure of the Topological Insulator $Bi_2Se_3$ Using Angle-Resolved Photoemission Spectroscopy: Evidence for Nearly Full Spin Polarization. *Phys. Rev. Lett.* **106**, 257004 (2011).

16. S. Souma *et al.*, Direct Measurement of the Out-of-Plane Spin Texture in the Dirac-Cone Surface State of a Topological Insulator. *Phys. Rev. Lett.* **106**, 216803 (2011).

17. C. Jozwiak *et al.*, Widespread spin polarization effects in photoemission from topological insulators. *Phys. Rev. B* **84**, *165113* (2011).

18. K. Okamoto *et al.*, Observation of a highly spin-polarized topological surface state in $GeBi_2Te_4$. *Phys. Rev. B* **86**, 195304 (2012).

19. D. Kong *et al.*, Ambipolar field effect in the ternary topological insulator $(Bi_xSb_{1-x})2Te_3$. By composition tuning. *Nature Nano.* **6**, 705 (2011).

20. J. G. Checkelsky *et al*., Bulk Band Gap and Surface State Conduction Observed in Voltage-Tuned Crystals of the Topological Insulator $Bi_2Se_3$. *Phys. Rev. Lett.* **106**, 196801 (2011).

21. Y. Wang *et al*., Gate-Controlled Surface Conduction in Na-Doped $Bi_2Te_3$ Topological insulator Nanoplates. *Nano Lett*. **12**, 1170-1175 (2012).

22. S-Y. Xu *et al*., Topological Phase Transition and Texture Inversion in a Tunable Topological Insulator $TlBiSe_2$. *Science*. **332**, 560-564 (2011).

23. T. Sato *et al*., Unexpected mass acquisition of Dirac fermions at the quantum phase transition of a topological insulator. *Nature Phys*. **7**, 840 (2011).

24. C. H. Park, S. G. Louie, Spin Polarization of Photoelectron From Topological insulators. *Phys. Rev. Lett*, **109**, 097601 (2012).

25. C. Jozwiak *et al*., Photoelectron Spin-flipping and texture manipulation in a topological insulator. *Nature Phys.* **9**, 293 (2013).

26. Supplementary materials.

27. L. Fu, Hexagonal Warping Effects in the Surface States of the Topological insulator $Bi_2Se_3$. *Phys. Rev. Lett.* **103**, 226801 (2009).

28. T. Zhang *et al*., Experimental Demonstration of Topological Surface States Protected by Time-Reversal Symmetry. *Phys. Rev. Lett.* **103**, 226803 (2009).

29. S. Tanuma, C. J. Powell, D. R. Penn, Calculations of Electron Inelastic Mean Free Paths. *Surf. Interf. Anal.* **21**, 165 (1994).

30. S. Muff *et al*., Separating the bulk and surface n- to p-type transition in the topological insulator $GeBi_{4-x}Sb_xTe_7$. *Physical Review B*. **88**, 035407 (2013).

31. Z. Ren *et al*., Fermi level tuning and a large activation gap achieved in the topological insulator $Bi_2Te_2Se$ by Sn doping. *Physical Review B*. **85**, 155301 (2012).

32. D. Hsieh *et al*., A tunable topological insulator in the spin helical Dirac transport regime. *Nature* **460**, 1101–1105 (2009).

33. Z. Ren *et al*., Optimizing $Bi_{1-x}Sb_xTe_{1-y}Se_y$ solid solution to approach the intrinsic topological regime. *Physical Review B*. **84**, 165311 (2011).

**Acknowledgments:**

We thank S. Zhu, T. Maegawa and K. Shibata for technical supports. The ARPES and SARPES measurements were performed at HiSOR, Hiroshima University with the approval of the Proposal Assessing Committee of HSRC (Proposal No.12-A-24, 11-B-12). The HAXPES experiment was performed at BL15XU of SPring-8 with the approval of NIMS Beamline Station (Proposal No. 2012B4908). This work was partly supported by KAKENHI (20340092), Grant-in-Aid for Scientific Research (B) of Japan Society for the Promotion of Science. K. K. and G. E. acknowledge support from the Japan Society for the Promotion of Science for Young Scientists.

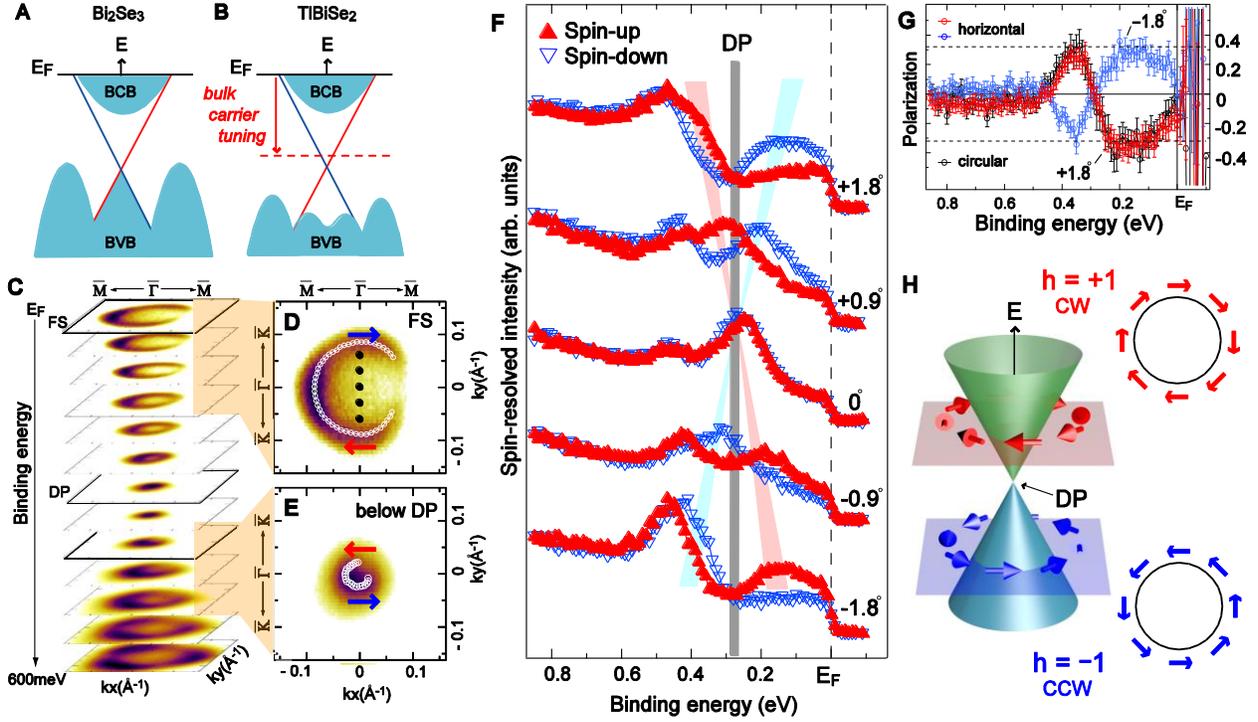

**Fig. 1**. **Topological surface state of TlBiSe$_2$ and its novel spin texture.** Schematic energy band structures for (**A**) Bi$_2$Se$_3$ and (**B**) TlBiSe$_2$, where shaded areas denote bulk continuum states including the bulk conduction band (BCB) and valence band (BVB) with a surface Dirac cone. The BCB is located at the $E_F$ in both grown samples, which leads to the large bulk conductions. These features have been obtained in previous works (6, 9, 10-12). (**C**) Constant energy contour maps at several $E_B$ from $E_F$ to 600 meV in 50 meV step. Selected constant energy contours at (**D**) $E_F$ (FS), (**E**) below the DP ($E_B$=450 meV). Intensity maxima obtained from the momentum distribution curves (MDCs) are denoted with open circles in the figures (**D**) and (**E**). The colored arrows indicate the spin orientations determined by SARPES measurement (see below). (**F**) Spin-resolved EDCs along Γ-K at the emission angles ($\theta$) from -1.8° to +1.8° in 0.9° step, which approximately correspond to the ***k*** positions denoted with closed circles in (**D**). Spin-up and spin-down spectra are plotted with closed upright (filled) and inverted (open) triangles. (**G**) Experimental spin polarizations taken after subtraction of a constant background as a function of $E_B$ for $\theta$=-1.8° and +1.8°. Maximum polarization of 32 % are denoted with dashed lines. Red and blue circles show the data taken with linear polarization and black ones denote that taken with circular polarization. (see main text and also supplementary information (26)). (**H**) Schematic picture of spin-helical surface Dirac cone with reversed spin helicity at the Dirac point (DP).

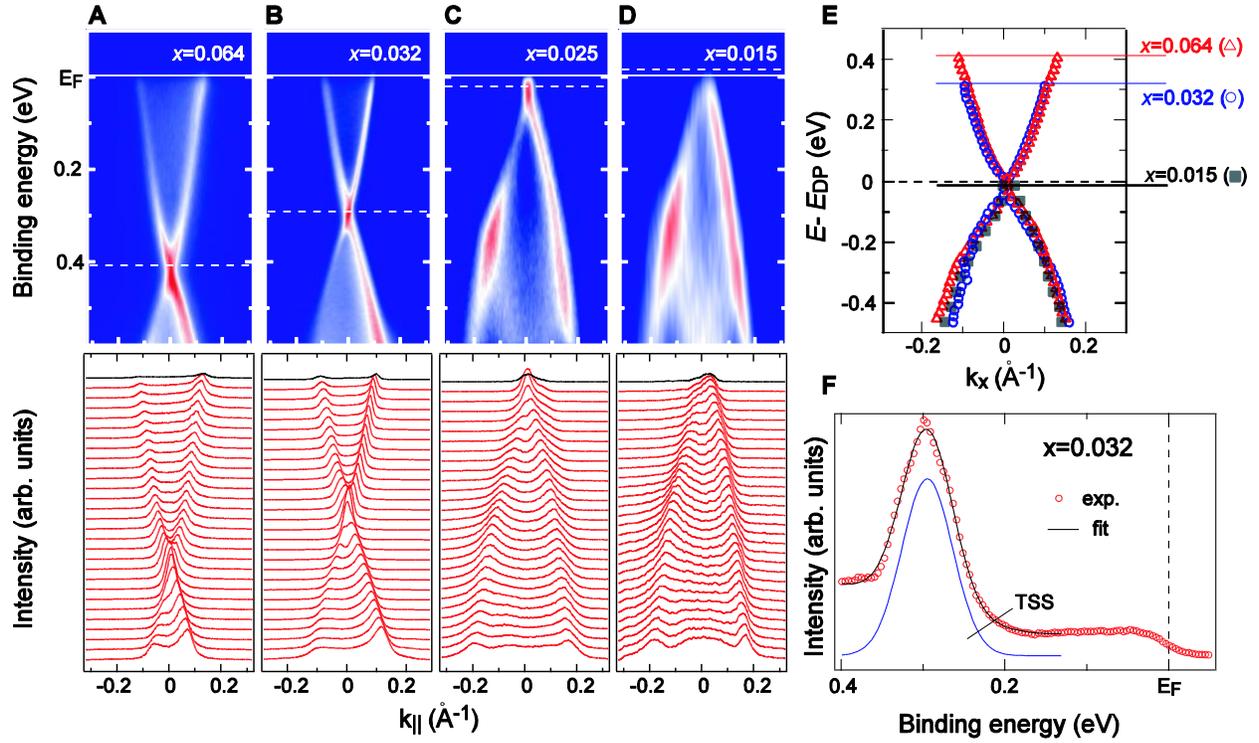

**Fig. 2. Evolution of a single surface Dirac cone with *x* in Tl$_{1-x}$Bi$_{1+x}$Se$_{2-\delta}$ (TBS):** (**A**)-(**D**) Measured band dispersions and momentum distribution curves (MDCs) for *x*=0.064-0.015 samples. Top row: image plots of band dispersions along the M-Γ-M direction obtained with *hν* = 17.5 eV. Bottom row: MDCs extracted from image plots in the top row. (**E**) Surface state dispersions for different *x* plotted with respect to the DP energy ($E_{DP}$). (**F**) EDC for TBS with *x*=0.032 at the central *k* position with a numerical fitting result (solid line).

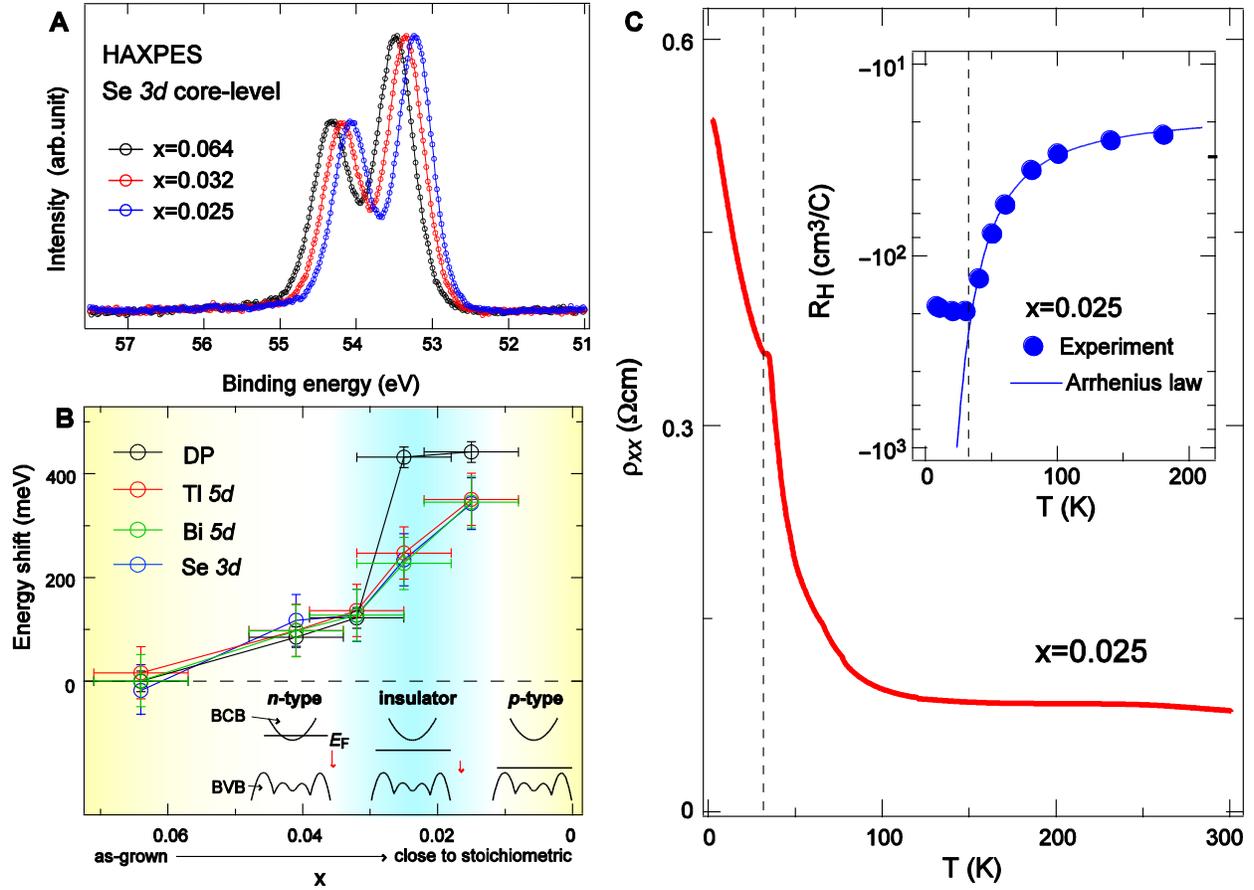

**Fig. 3. Evolution of deeply buried bulk state with *x* in Tl$_{1-x}$Bi$_{1+x}$Se$_{2-\delta}$ obtained by hard x-ray photoemission spectroscopy.** (**A**) Se *3d* core-level with spin-orbit splitting for TBS with different *x*. The samples with *x*=0.064, 0.032, 0.025 are denoted with black, red and blue open circles and thin solid lines. (**B**) Comparison of *x* dependence with respect to the sample with *x*=0.064 for bulk core-levels and surface DP energy ($E_{DP}$) in Fig. 3**A**. The energy shifts for the Tl *5d*, Bi *5d* and Se *3d* core-levels are denoted with red, green and blue colored circles and solid lines, respectively. Inset figures in panel **B** schematically show the relationship between the deeper-lying bulk state and chemical potential. (**C**) Temperature dependence of longitudinal resistivity ($\rho_{xx}$) for *x*=0.025. Inset shows the observed temperature dependence of the Hall coefficient ($R_H$) with numerical fitting results assuming the Arrhenius low (solid line)

**Supplementary Materials:**

Supplementary Text

Figures S1-S7

References

**Supplementary Materials:**

**This file includes:**

<u>SI. 1</u>  **Sample growth and preparations**
<u>SI. 2</u>  **Evaluations of chemical compositions by EPMA analysis**
<u>SI. 3</u>  **Experimental conditions**
<u>SI. 4</u>  **Evolution of surface band structure with time**
<u>SI. 5</u>  **Details of spin- and angle-resolved photoemission spectroscopy**
<u>SI. 6</u>  **Estimation of electronic structures deep in the bulk**

## SI. 1 Sample growth and preparations

Single crystals of $Tl_{1-x}Bi_{1+x}Se_{2-\delta}$ ($x$=0.015-0.064) were grown by a standard procedure using Bridgeman method with high purity elements (Bi, Se; 99.999%, Tl; 99.99%). First, the temperature of the sample was raised above the melting point around 800 °C, and kept for two days to improve a homogeneity. Then, the sample was cooled down to 100 °C very slowly for twenty days. The quality of all samples was checked with Laue diffraction, X-ray diffraction. The $x$ values were determined by electron probe micro-analysis (EPMA) (see also SI 2.). The carrier-type of samples was evaluated from the measurement of Seebeck coefficients. After all the above-mentioned procedures were completed, a part of ingot was cut into a suitable size of 2 ×2×0.5 mm$^3$ for the photoemission experiments. For transport measurements, the standard Hall-bar set-up was made on the cleaved surface with a thickness of 0.46 mm and Au wires were attached with silver paint.

## SI. 2 Evaluations of chemical compositions by EPMA analysis

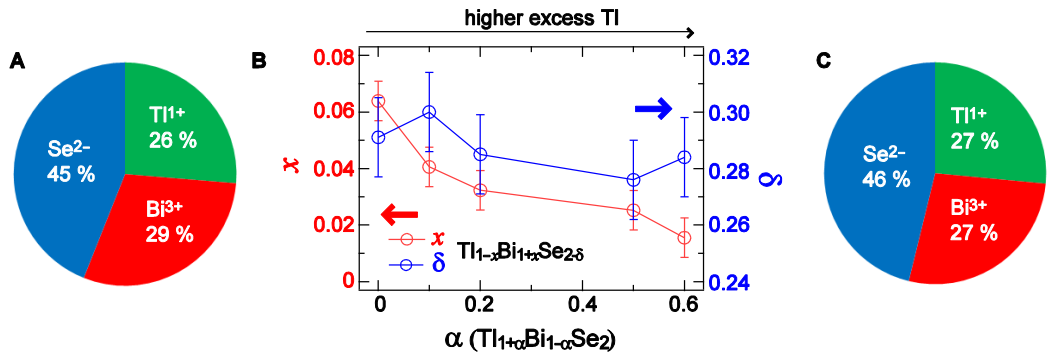

**FIG. S1:** Estimated chemical compositions by EPMA analysis. **A**, Result of chemical composition analysis for $\alpha$=0. **B**, Evaluated $x$ values versus $\alpha$ values. **C**, Result of chemical composition analysis for excess Tl sample ($\alpha$=0.6).

In order to understand the origin of spontaneous electron doping effect for the sample of TlBiSe$_2$, EPMA analysis has been conducted at N-BARD in Hiroshima University (JEOL JXA-8200). The result for the sample fabricated starting with the composition ratio of Tl (1): Bi (1): Se (2) is summarized as a circle graph in Fig. S1A. It is noticed that the estimated compositions deviate from the stoichiometric values, that is, (i) the Se concentration is less than 50% and (ii) the ratio of Tl concentration to that of Bi is less than unity. First, the item (i) tells us that the chalcogen-site vacancies are present in the synthesized sample, and a portion of Tl and Bi ions may also be substituted at chalcogen sites (anti-site defects). The chalcogen vacancies often take place in chalcogenide systems that act as donors while the anti-site defects behave as accepters. Note that, anti-site defects could be a minor effect because the electronegativities of Tl and Bi are smaller than that of Se. The item (ii) has never been reported so far for TlBiSe$_2$, which might be considered to stem from substitutional-type defects at cation sites, where a small portion of Tl$^{1+}$ sites are replaced by Bi$^{3+}$. By taking into account the number of valence electrons of both ions, one can easily find that this type of defect works as an electron donor as expressed with the following equation,

$$Bi + Tl^{1+} = Bi^{3+} + Tl + 2e^- \quad \text{(eq. SI2-1).}$$

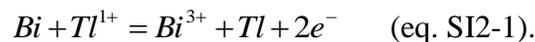

Therefore, we need to describe the actual chemical composition of samples as $Tl_{1-x}Bi_{1+x}Se_{2-\delta}$, where $x$ and $\delta$ denote the amount of cation-site defects and anion-site defects, respectively. In order to conduct hole doping in $TlBiSe_2$, it is necessary to reduce $x$ down to nearly a stoichiometric value. In this work, we have tried to reduce the substitutional defects at the cation-sites by introducing excess Tl as an initial composition $Tl_{1+\alpha}Bi_{1-\alpha}Se_2$, where $\alpha$ denotes the amount of excess Tl atoms and the amount of Se is fixed in the sample growth procedure. Figure S1B shows the $x$ and $\delta$ values determined by the EPMA analysis as a function of $\alpha$, where the error bars come from the statistical error in the EPMA analysis at several parts of the sample. Here, one can find that $x$ decreases with increasing $\alpha$ while $\delta$ tends to be constant. For $\alpha$=0.6, the Bi/Tl ratio is close to unity as shown in Fig. S1C. It is thus shown that a nearly stoichiometric composition can be achieved.

## SI. 3 Experimental conditions

Angle-resolved photoemission spectroscopy (ARPES) and spin- and angle-resolved photoemission spectroscopy (SARPES) experiments were performed with synchrotron radiation at the ESPRESSO end station attached to the APPLE II variable polarization undulator beamline (BL9B) (34) and the high-resolution ARPES end-station BL9A of Hiroshima Synchrotron Radiation Center (HiSOR). Each photoemission spectrum was acquired with the hemispherical photoelectron analyzer (VG-SCIENTA R4000) at 40 K for SARPES measurement and at 80 K for ARPES. The energy and angular resolutions were set below 20 meV and 0.1 ° for ARPES, and 30 meV and 0.7 ° for SARPES experiment, respectively. The photon energy ($h\nu$) was tuned at 17.5eV and 17.7eV for ARPES and SARPES measurements, respectively. Clean surfaces were obtained by cleaving in ultrahigh vacuum at low temperature. The hard X-ray photoelectron spectroscopy (HAXPES) measurements were performed at BL15XU of SPring-8 ($h\nu$ = 5.95 keV). The total energy resolution was set to 240 meV. All the HAXPES experiments were performed at room temperature for the same sample sets as used in the ARPES and SARPES experiments without cleaving the samples.

## SI. 4 Time evolution of surface band structure

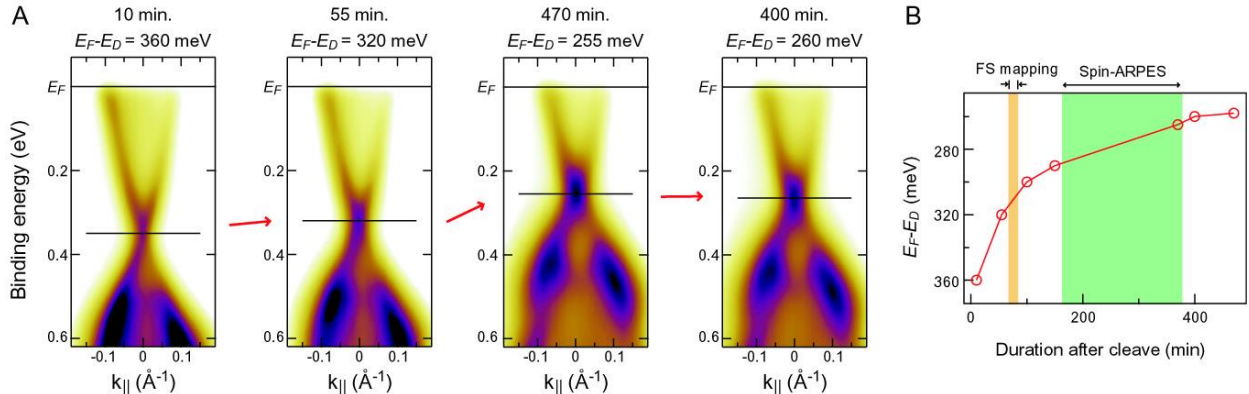

**FIG. S2: Time evolution of surface state for as-grown sample. A,** ARPES maps acquired at 10, 55, 400, 470 min after the sample cleaving. **B,** Time dependent energy shift of the Dirac point (DP). Red and green rectangles denote the time window for the Fermi surface mapping and the SARPES measurement.

As the fresh surfaces for all measurements were obtained by *in-situ* cleaving the (111) plane of TlBiSe$_2$ samples at ~40 K, residual gases adsorbed on the sample surface can gradually introduce hole-type carriers at the surface, causing a downward-shift of $E_F$. To check whether this energy shift affect the Fermi surface mapping and the SARPES measurement, the evolution of Dirac point (DP) energy after cleaving are plotted as a function of time. After 7 hours, the energy shift of the DP by hole-doping was saturated at 260 meV, where the bulk conduction band still crosses $E_F$. For the Fermi surface mapping, it nearly takes 20 min, causing 10 meV energy shift soon after cleaving. Thus, to complete the Fermi surface mapping, this energy shift is ignorable that is smaller than the energy resolution. On the other hand, the SARPES measurement started after the energy shift was saturated because it takes comparatively longer time to obtain sufficient statistics of spin-resolved data (see also SI. 4).

**SI. 5  Detailed description of spin- and angle-resolved photoemission spectroscopy**
**5-1. Very low energy electron diffraction (VLEED) type spin detector**

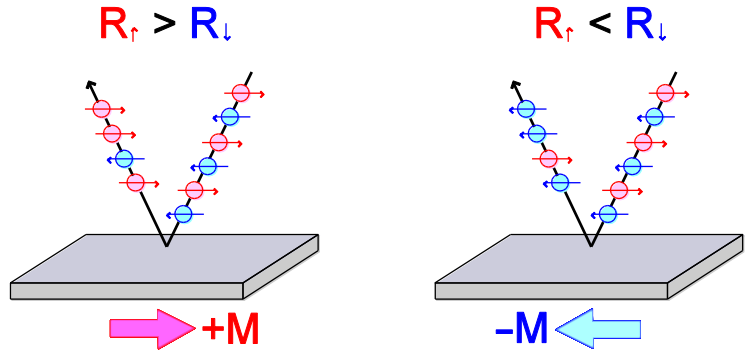

**FIG. S3: Spin dependent reflection at magnetized target for very low energy electron diffraction type spin detector.**

Electron spin detectors are based on spin-dependent scattering events leading to asymmetric intensities of scattered electrons. The unknown spin polarization of incident electron beam along a given axis, ***P*** is proportional to the normalized intensity difference, *A*, between two distinct scattering channels, $I_1$ and $I_2$. The proportionality is generally expressed as the effective Sherman function, $S_{eff}$, which evaluates the spin analyzing power of the given scattering process and must depend on the target of electron spin detectors. Finally, the spin polarization of incident electron beam is determined by the following equation,

$$P = \frac{A}{S_{eff}} = \frac{1}{S_{eff}} \frac{I_1 - I_2}{I_1 + I_2} \qquad \text{(eq. SI5-1)}.$$

Due to the loss of electron intensity through the scattering process, the spin-resolved experiment is usually performed with increased analyzer slit width, which sacrifices the energy and momentum resolutions. Nevertheless, long counting scans are still required to achieve a high S/N ratio. In order to argue the efficiency of a given spin detector, one often defines the "figure of merit", ε, given by following equation,

$$\varepsilon = \frac{I}{I_0} S_{eff}^2 \quad \text{(eq. SI5-2)}.$$

where $I_0$ and $I$ are total incident electrons and total counted electrons (sum up $I_1$ and $I_2$), respectively. The statistical error of a spin polarization measurement is given by:

$$\frac{\delta P}{P} = \frac{1}{S_{eff}} \frac{\delta A}{A} = \frac{1}{\sqrt{\varepsilon I_0}} \quad \text{(eq. SI5-3)}.$$

To overcome this problem and to realize a highly efficient measurement of the spin polarization for very steep energy dispersion such as surface Dirac cones, we utilized very low energy electron diffraction (VLEED) type electron spin detector as shown in Fig. S3. This technique is based on the spin-dependent electron reflectivity of very low energy electrons, which originates from the exchange-split band structures of the ferromagnetic target. Because of higher scattering probability of the low-energy electrons ($E_k \sim$ 6-10 eV) with respect to the high-energy electrons ($E_k \sim$ 20-100 keV) for the Mott scattering and relatively higher effective Sherman function, the VLEED-type spin detector presents 100 times higher $\varepsilon$ than that for the Mott-type conventional spin detector. For the VLEED-type spin detector, the intensity asymmetry is re-defined by the following equation,

$$A_\alpha = \frac{I_{\alpha+} - I_{\alpha-}}{I_{\alpha+} + I_{\alpha-}} \quad \text{(eq. SI5-4)}$$

where $I_{\alpha+}$ ($I_{\alpha-}$) is the reflected electron intensity by the positively (negatively) magnetized ferromagnetic target along the direction $\alpha$. Thus, the electron spin polarization along the quantization axis is described with the following equation.

$$P_\alpha = \frac{A_\alpha}{S_{eff}} \quad \text{(eq. SI5-5)}$$

## 5-2. Experimental set-up of "ESPRESSO" machine at HiSOR

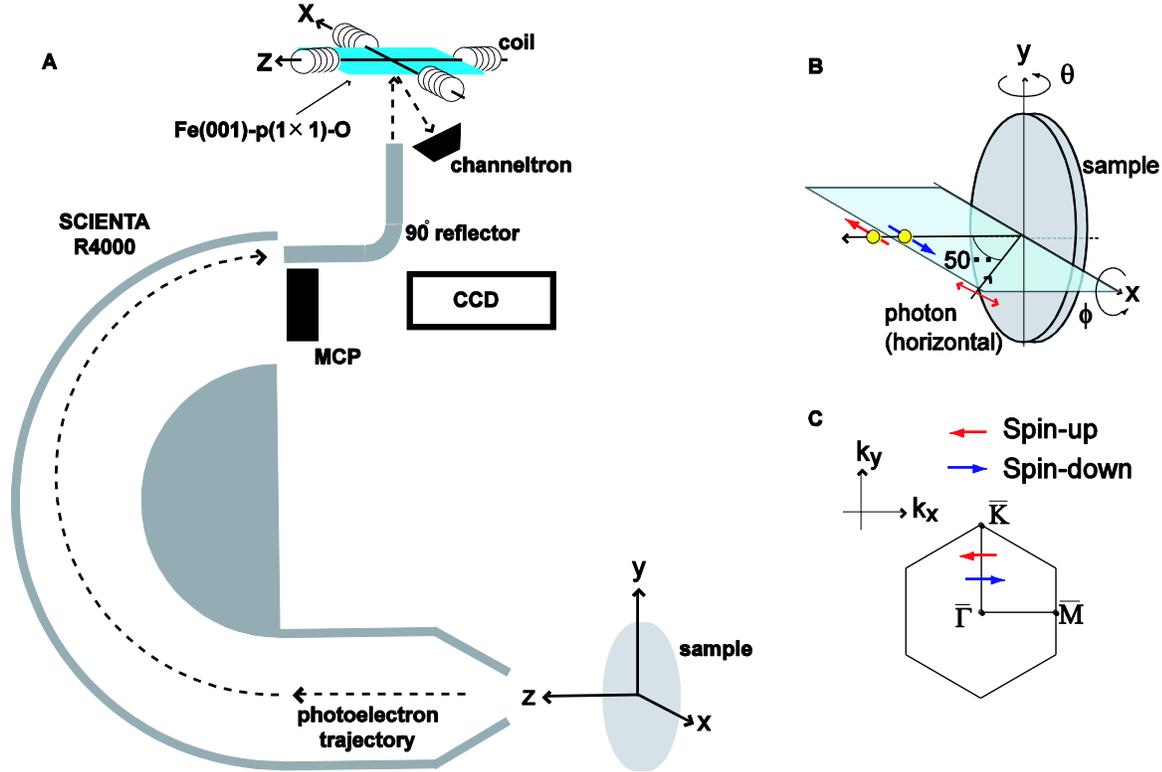

**FIG. S4: Experimental set-up: A**, Schematic design of the ESPRESSO machine, where the spin detector is connected via 90° electron deflector to the high-resolution hemispherical electron analyzer (VG-SCIENTA R4000). **B**, Schematic illustration of experimental geometries for ARPES and SARPES measurement using horizontally polarized synchrotron radiation. **C**, Measureable spin directions with surface Brillouin zone.

The ESPRESSO system is composed of the hemispherical analyzer (SCIENTA R4000) and the VLEED spin detector as shown in Fig. S4A. Our VLEED spin detector consists of a ferromagnetic target and two pairs of coils for two orthogonal magnetization directions along X and Z axes at the target, which enables us to measure both in-plane and out-of-plane spin polarizations in the normal emission configuration. It has achieved efficiency of $\varepsilon \sim 1.1 \times 10^{-2}$, which is about 100 times higher than that of the conventional Mott-type spin detector (34). The photoelectron was excited with various polarized synchrotron radiation. Figure S4B shows the details of experimental set-up and the measured spin polarization vectors are parallel or antiparallel to the Γ-M line of the surface Brillouin zone in Fig. S4C.

## 5-3. Detailed description of spin-resolved photoemission analysis with the ESPRESSO machine

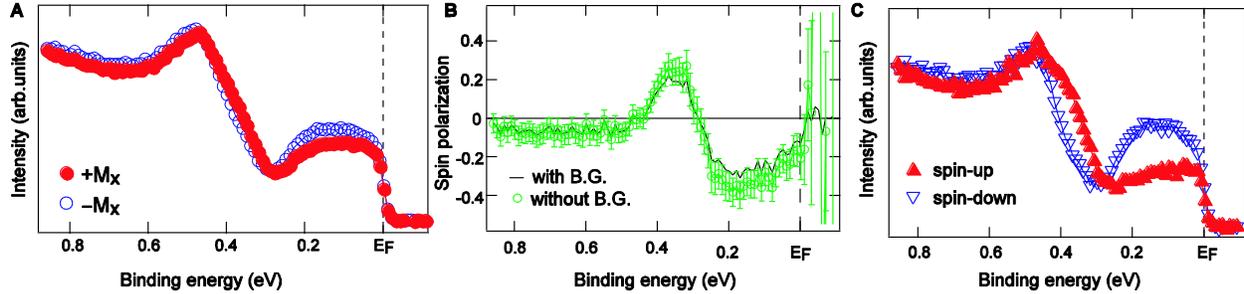

**FIG. S5**: **A**, Energy distribution curves (EDCs) measured with different magnetization directions that are parallel (red closed circles) and anti-parallel (blue open circles) to $X$ axis. **B**, Spin polarizations as a function of binding energy obtained from the spin-resolved analysis (see main text). **C**, Finally obtained EDCs in spin-up (red filled triangles) and spin-down (blue open triangles) channels.

Here, we show how to obtain the spin resolved spectra with the ESPRESSO machine at HiSOR. Without top-up electron injection into a storage ring of HiSOR, the photoelectron intensity superimposed on the extrinsic background is reduced with unavoidable decay of the electron current in the storage ring. This could be a problem because the photoelectrons need to be counted at least twice with opposite in-plane magnetization directions of the target to get the intensity asymmetry (see eq. SI5-4). In order to solve this problem, we have taken the SARPES spectra in a measurement cycle; $+M \to -M \to -M \to +M$ and the two spectra are finally averaged for each magnetization direction as shown in Fig. S5A. Typically, it takes approximately 20 minutes to complete a cycle, which corresponds to the time scale for the band shift of ~2 meV (see SI. 4). Therefore, this shift is negligible with respect to the energy resolution. To achieve sufficient statistics of the spectra, we repeat the same procedure several times and subsequently the measured spin polarization spectra are averaged as shown in Fig. S5B.

Focusing on the intensity above $E_F$, we find non-spin polarized background signal probably owing to the higher-order light from the monochromator. This unpolarized background can mask the spin polarization of photoelectron emitted by the first-order light, and should be removed before the spin analysis. In Fig. S5B, the green open circle shows the spin polarization after the background subtraction, which is assumed to be constant. Comparing to the spin polarization without the background subtraction, one can see slightly larger spin polarizations. The resultant spin-resolved EDCs are shown in Fig. S5C.

## 5-4. Experimental verification of spin polarized topological surface state

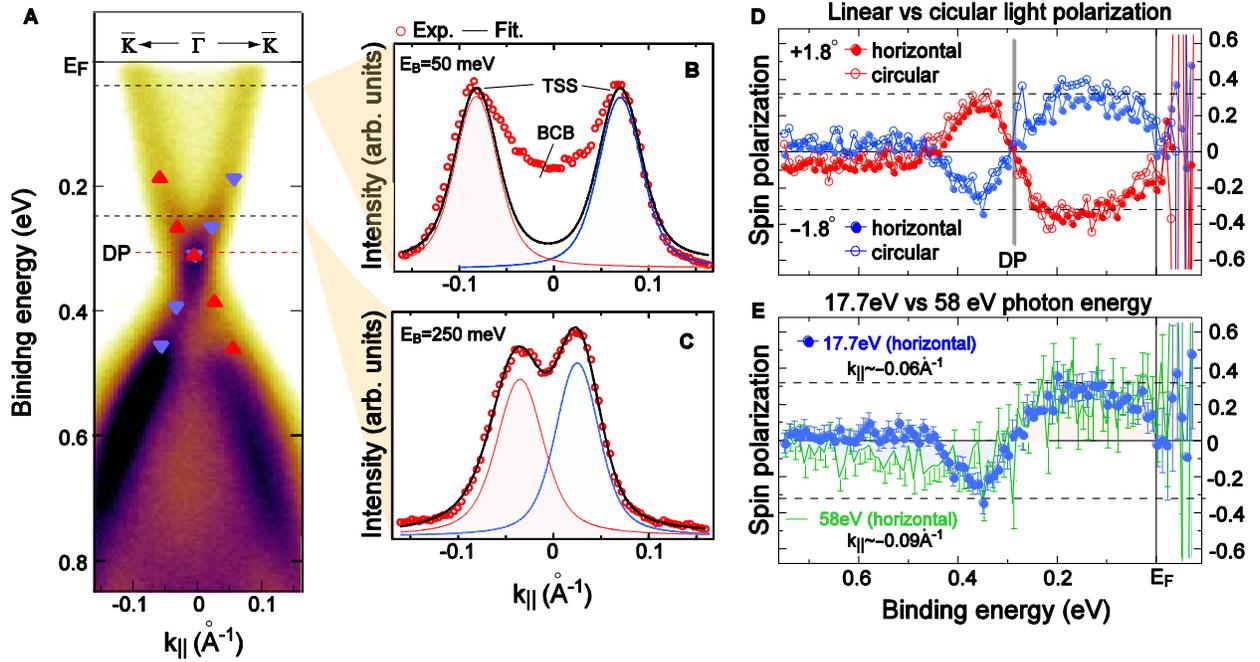

**FIG. S6**: **Observation of spin helical topological surface state with variable light polarizations and photon energies.** **A**, ARPES intensity map of topological surface state with DPs at $E_B = 310$ meV. **B** and **C** show momentum distribution curves at several $E_B$ above DP (**B**: near $E_F$, **C**: $E_B = 250$ meV). **D**, Spin polarizations taken with horizontally and circularly polarized lights. Red and blue circles indicate the spin polarizations obtained by (closed) horizontally and (open) circularly polarized lights at +1.8 ° and -1.8 °, respectively. **E**, A comparison of spin polarizations taken with different photon energies of 58 eV (green marks) and 17.7 eV (blue marks). Solid lines in **D** and **E** denote the size of typical spin polarization 32%.

In order to determine the spin-helical texture of a surface Dirac cone exhibiting very steep energy dispersions, high energy and angular resolutions are required for the SARPES measurement. In particular, the spin polarization could be reduced extrinsically due to the photoelectron intensity from the un-polarized bulk states that are very close to the surface state. Therefore, the $h\nu$ was tuned to suppress the bulk intensity with respect to the surface Dirac cone. Figure S6A shows the ARPES intensity map acquired ($h\nu = 17.7$ eV), accompanying an X-shaped surface Dirac cone and also an *M*-shaped bulk state below the DP. To confirm bulk contributions to total photoelectron intensity, momentum distribution curves (MDCs) at several binding energies as depicted in Fig. S6B and C. Near the DP in the bulk band gap region, the experimental data can be fit well with two Lorentzian functions, which tells us that there is only a surface Dirac cone at this binding energy and the bulk contribution to total photoelectron counts is negligible. On the other hand, additional photoelectron intensities appear between two branches of surface Dirac cone near $E_F$, which indicates the photoelectron intensities from bulk states are still visible with this $h\nu$. Thus, in order to avoid extrinsic factor that reduces the apparent spin polarization, we have tried to see the spin polarizations of surface Dirac cone within the bulk band gap. Peak positions determined by spin-resolved EDC for each spin directions are plotted in Fig. S6A (see Fig. 1F in the main text). And then, representative spin polarizations are shown in Fig. S6D. One can find an anti-symmetric feature of spin polarizations that changes their signs at the Γ point and DP. The observed spin polarization features are characteristic of spin-helical Dirac cone. Next, in order to confirm if the observed spin polarizations originate from the initial state of the surface state, the same measurements with different light polarizations and $h\nu$ are performed. The observed spin polarization with circularly polarized light is shown in Fig.S6D. Although there are slight differences in the spectra taken with linearly polarized light, the characteristic feature in spin polarizations with a sign reversal at the DP is again observed. Moreover, this feature does not strongly depend on $h\nu$. We thus conclude that the experimental spin polarizations originate from the initial-state spin helical textures at the surface of TlBiSe$_2$.

### SI. 6 Estimation of electronic structures with $x$ dependences

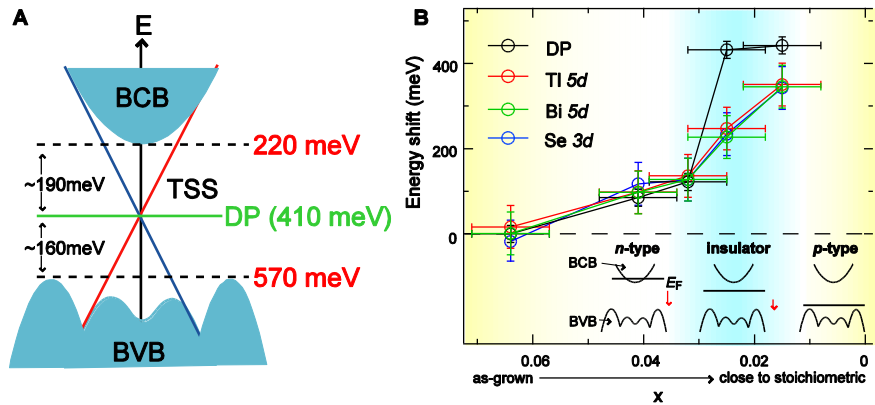

**FIG. S7: Energy diagram of deep bulk and surface electronic structures. A**, Schematics of energy band structure for bulk and surface states around Γ point revealed by the previous ARPES experiment (10). **B**, Energy shift of Tl *5d*, Bi *5d* and Se *3d* core-levels relative to that for $x$=0.064 as a function of $x$. Inset figures show schematic images of bulk state (BCB: bulk conduction band, BVB: bulk valence band) and chemical potential.

Figure S7A shows schematic energy band structure near Γ point for TlBiSe$_2$, where the shaded area represents the bulk continuum states together with the surface Dirac cone. This characteristic energy dispersion has been determined by the previous work of ARPES (10) using vacuum ultra-violet light (VUV), which is considered to be a surface sensitive probe. Since the bottom of bulk conduction band (BCB) bottom is known to be separated by 190 meV from the DP at Γ point, the BCB near the surface and the DP are determined to be located around $E_B$ = 220 meV and 410 meV for $x$ = 0.064, respectively. The observed core-level shifts as a function of $x$ with respect to that for $x$=0.064 are summarized in Fig. S7B together with the DP energy shift obtained by the ARPES measurement (Fig. 2 in the main text). We note that the core-level shifts nearly coincide with those of the DP above $x$=0.03 and the DP energy shift starts to deviate from those of the bulk core level shifts at $x$=0.025, where the BCB bottom is located within the energy band gap. Such a discrepancy can be explained by the band bending formation as a result of the bulk insulating property.

### Reference:

34. Okuda, T. *et al*. Efficient spin resolved spectroscopy observation machine at Hiroshima Synchrotron Radiation Center. *Rev. Sci. Instrum*. **82**, 103302 (2011).